# Electrochemical synthesis of superconducting $MgB_2$ thin films: a novel potential technique

S H Pawar*, A B Jadhav, P M Shirage and D D Shivagan

Superconductivity Lab.,
School of Energy Studies,
Department of Physics,
Shivaji University, Kolhapur- 416 004 INDIA.

**Abstract**

A complexing molecule electrodeposition technique has been developed for the deposition of low cost, lightweight $MgB_2$ films, for the first time. Different deposition parameters such as bath composition, deposition potential, current density, deposition time were studied and optimized to give uniform, homogeneous and sticky films. The $MgB_2$ films were deposited at a constant potential of -1.3 V with respect to Saturated Calomel Electrode (SCE) onto silver substrate from aqueous bath and at -3.2 V with respect to SCE electrode onto silver substrate from non-aqueous bath. XRD, SEM and EDAX techniques are used to characterize these films. The films formed from non-aqueous electrochemical bath show the superconducting transition temperature at $T_c$ = 36.4 K. The method developed is of less energy inputs than any other existing methods and of versatile nature having potential for large scale applications of $MgB_2$ superconductors.

*Author for correspondence

*E-mail: pawar_s_h@yahoo.com*

# I. INTRODUCTION

The recent discovery of superconductivity in the binary intermetallic compound magnesium diboride at $T_c$ = 39 K by J. Akimitsu [1] has created great interest amongst many researchers. This reported $T_c$ value is much higher than that of the highest $T_c$ reported for any other intermetallic compounds such as $Nb_3Ge$, $T_c$ = 23.2 K [2] and borocarbides [3] as well as for any non-oxide and non-cuprate based materials. Further, this value of $T_c$ is above the limit suggested by the BCS theory, which raises the question, whether this high $T_c$ in $MgB_2$ is due to electron -phonon coupling or other mechanism?

$MgB_2$ has a simple hexagonal crystal structure with P6/mmm symmetry in which graphite like honeycomb layer of boron atoms is stacked in between two hexagonal layers of magnesium atoms. Even though superconductivity in $MgB_2$ is much lower than high temperature cuprates it is important because of its superiority over high $T_c$ materials. Firstly, it is made up of light and cheap elements abundantly available in nature. Secondly, unlike cuprates $MgB_2$ is good metal where there is no high contact resistance between grain boundaries eliminating weak link problem that has plagued wide spread commercialization of the cuprates. Further, the conduction electron density and normal state conductivity are one to two orders of magnitude higher for $MgB_2$ than for the cuprates in the present day wires and thin films. These features make $MgB_2$ attractive for many applications.

After the discovery of superconductivity in $MgB_2$, several research groups have made the attempts to synthesize high quality samples of $MgB_2$ in the forms of wires [4], tapes [5] and films [6]. Thin films of $MgB_2$ have been prepared by pulse laser deposition (PLD) technique [6] and electron beam evaporation technique [7] requires post-annealing process. These films are prepared by using pure bulk $MgB_2$ sample under high vacuum pressure with high temperature. As such overall processes of film preparation are lengthy and costly. The authors have done the pioneering work in electrochemical processing of high temperature superconductors (HTSC) and have successfully demonstrated the complexing molecule electrodeposition of superconducting thin films, wires and pipes with any geometry of most of Bi-Sr-Ca-CuO [7-9], Dy- Ba-CuO [10], Y-Ba-CuO [11,12], Tl-Ba-Ca-CuO [13] and Hg-Ba-Ca-CuO [14] based HTSC systems. Electrosynthesis provides highly reactive mixture on atomic scale that markedly reduces the time and temperature. Also it gives stoichiometric, highly conducting, non-pours and fine-grained deposit. Hence in the present investigation complexing molecule electrodeposition technique has been developed to obtain $MgB_2$ thin films.

In the present paper we are reporting, for the first time, the complexing molecule electrodeposition of $MgB_2$ thin films onto silver substrate from both aqueous and non-aqueous baths. The different electrochemical deposition parameters such as bath composition, deposition potential, current density and deposition time are optimized to get good quality stoichiometric $MgB_2$ films. These thin films are characterized by X-ray diffraction, SEM, EDAX and electrical properties.

## II. EXPERIMENTAL

The complexing molecule electrodeposition bath was prepared by dissolving 50 mM magnesium acetate and 50 mM boric acid in double distilled water as aqueous bath and in dimethyl sulphoxide (DMSO) as non-aqueous bath. The quantity of the constituents in the electrodeposition bath was adjusted to get stoichiometric $MgB_2$ films. The electrodeposition cell consists of a three-electrode system with saturated calomel electrode (SCE) as a reference electrode, metal substrate (silver and stainless steel foil) as working electrode and graphite as a counter electrode. The MgB$_2$ films were deposited using scanning potentiostat/galvanostat (EG and G model 362) under potentiostatic conditions at room temperature. An Omnigraphic X-Y recorder (model 2000) was used to record the voltammograms and hence to determine the deposition potentials. The as-deposited films were then heated at moderate temperatures of about 450ºC for 5 hr in a sealed tube furnace.

X-ray diffraction data were obtained with a microcomputer controlled Philips PW-3710 diffractometer using CuK$_\alpha$ radiation. The Scanning electron micrographs were carried out using Philips model provided with the energy dispersive X-ray spectroscopy (EDAX) probe for the elemental quantitative analysis. The samples were cooled in Helium close cycle refrigerator (HC-2D Model). The standard four-probe method was employed to measure the resistivity and hence to determine the superconducting transition temperature $T_c$.

### III. RESULTS AND DISCUSSION

The electrochemical studies of Boron show that it can be deposited as per the following reactions [15],

$H_3BO_3 + 3H^+ + 3e^- \rightarrow$ B + 3H$_2$O at -0.86 V vs. NHE from aqueous bath, (1)

and

$B^{3+} + 3e^- \rightarrow$ B at - 2.29 V vs. NHE from non-aqueous bath (2)

The electrochemical deposition of elemental boron has limitations as it easily forms boro-hydride, *in-situ* at higher potentials, which is unstable at room temperature when exposed to the atmosphere. This is given by the following reaction,

$H_3BO_3 + 4H^+ + 4e^- \rightarrow$ BH + 3H$_2$O at -1.873 V vs. NHE for aqueous bath (3)

In case of Magnesium, it can be deposited as [16],

$Mg^{2+} + 2e^- \rightarrow Mg$     at -2.36 V vs. NHE from aqueous bath,     (4)

and

$Mg^{2+} + 2e^- \rightarrow Mg$     at -1.1 to -2.36 V vs. NHE from non-aqueous bath     (5)

Though the electrodeposition of magnesium is easy, it is found to be sensitive to oxygen and it becomes milky-white magnesium oxide layer when dried in the open atmosphere.

The different values of deposition potentials for Mg and B, mentioned in above reactions show that a simple electrochemical bath will not be suitable to yield $MgB_2$ films. The formation of magnesium diboride films can be made possible by three different ways using electrochemical technique: (1) boron can be intercalated in magnesium films, (2) boron and magnesium can be co-deposited simultaneously, (3) A complexing molecule of magnesium and boron ($Mg^{++} + B^{+++}$) can be formed and electrochemically decomposed at a fixed potential.

Figure 1 shows the polarization curves for the deposition of magnesium-boron film from a complexing molecule bath consisting of 13 ml magnesium acetate (50 mM) and 22 ml boric acid (50 mM) solutions made in aqueous and non-aqueous solvents. In case of aqueous bath it is seen that the current increases at -1.3 V with respect to SCE (plot a). It is due to the dissociation of complex molecule into Mg and B cations at the anode, which then migrate towards Ag substrate and get deposited. It has been seen that there is a steep rise in current beyond -1.3 V with respect to SCE with increase in potential and observed the gas evolution. The gas evolution may be due to the decomposition of water molecule in aqueous bath as well as borohydride. In case of non-aqueous complexing bath (plot b) it is seen that the current increases at the potential of -3.2 V vs SCE without gas evolution. This deposition potential is comparatively higher than the aqueous bath.

The figure 2 shows the variation of deposition current density with deposition time for the magnesium-boron films onto silver substrate from both aqueous bath (a) and non-aqueous bath (b). From the graph it is seen that current density suddenly decreases in first few seconds and remains constant. This sudden decrease may be due to the formation of double layer at the electrode-electrolyte interface. It is further noted that the current slightly increases with deposition time. Which may be due to the discharge of double layer at electrode-electrolyte interface. The current density is found to be greater for the aqueous bath (0.80 mA/cm$^2$) than for non-aqueous bath (0.35 mA/cm$^2$). The film thickness depends on the cell current and time of deposition, while the deposition rate depends on the current density of cell. The optimum thickness of 1.5 μm is obtained for 30 min. deposition from aqueous bath while 1 μm thickness is achieved for 45 min. deposition from non-aqueous bath. It is observed that as-deposited films from aqueous bath are gray-white, while the films from non-aqueous bath are gray in colour.

The electrodeposited films from both aqueous and non-aqueous baths are needed to give post deposition heat treatments to yield superconducting property in MgB$_2$ films. The films deposited from non-aqueous bath gives the uniform and dense brownish-black films after moderate heating at 450 $^o$C in closed furnace. However, the films from aqueous bath turns to be whitish-yellow in colour. The films were also deposited onto stainless steel substrate from non-aqueous bath, keeping all preparative parameters same. However these films are found to be non-uniform.

The XRD pattern of the MgB$_2$ film deposited onto silver substrate from aqueous bath is shown in figure 3 (a). The pattern shows the phases of Mg and MgO along with MgB$_2$. The MgB$_2$ peaks are indexed with the hexagonal indices [1, 17], of which the space group is P6/mmm with cell parameters a = 3.086 Å, c = 3.524 Å. The appearance of MgO peaks in MgB$_2$ thin film may be due to the air oxidation of Mg or may be electro-oxidation. To avoid the electro-oxidation from aqueous bath we have selected non-aqueous bath formed using dimethyl sulphoxide (DMSO) as a solvent because it has high dielectric constant and prevents oxidation reaction.

Figure 3(b) shows the XRD pattern for the deposition of magnesium boron film from non-aqueous bath. It is seen that all peaks are of $MgB_2$ and no major impurity phases are observed. The lattice parameters are calculated, and found to be a = 3.025 Å and c = 3.429 Å. They are in good agreement with the standard results [17].

The XRD pattern of the $MgB_2$ film deposited onto stainless steel is shown in figure 3(c). It is seen that the films have the mixed phases of Mg, MgO and $MgB_2$ phase.

In $MgB_2$ superconductors the BCS conduction mechanism is similar in a, b and c-directions. As a result it does not show anisotropic property [18] like cuprate superconductors. Hence the orientation of the films do not affect the conduction properties.

Figure 4(a) shows the SEM of the film deposited onto Ag substrate from aqueous bath. It can be seen that the film is not homogeneous and the oxide phases are clearly observed onto the upper surface layer. However, the granulars with bigger size of $MgB_2$ are seen behind this layer, as marked on the microphotograph. Hence it could be possible to get $MgB_2$ rich films if one can maintain the inert environment to avoid oxidation. EDAX pattern shows the presence of Magnesium and Boron elements in the film. However, due to the contribution of oxide species (20.88 At %) and Ag substrate it is difficult to estimate the elemental ratio of Mg:B.

The SEM microphotograph of the film deposited onto Ag substrate from non-aqueous is shown in fig. 4 (b). It is seen that film is uniform, pore free, dense and very fine grained metallic deposit. No oxide species are distinctly observed on SEM. However, the EDAX pattern shows the minor quantity of oxygen (4.13 %) atoms in the deposits. Deleting the contribution of the Ag substrate, the ratio of Mg:B is found to be 1.2:1.9. This is in close agreement with the ratio of 1:2; with the consideration that some of the magnesium atoms are associated with oxygen to form MgO, quantitatively very small, as it could not be evident in the XRD pattern.

The temperature dependence of the resistivity of these $MgB_2$ films is shown in figure 5. It is seen that the resistance of the sample deposited from aqueous bath doesn't attend superconducting state upto 10 K. However, the film deposited from the non-aqueous bath shows metallic nature and sharp superconducting transition is observed at 36.4 K. The present sample contains some amount of oxygen, as observed in EDAX. However it has not considerably degraded the transition temperature $T_c$. On the contrary the small traces of oxygen has helped to improve the critical current and the superconductivity of the sample in stronger magnetic field as reported by Ecom *et al.* at University of Wisconsin-Madison and Princeton [19].

We have attempted to understand the mechanism of the formation of $MgB_2$ thin films by electrodeposition technique. Earlier some workers have deposited the thin films of $MgB_2$ by the process of diffusion. The precursor films of Boron have been taken and magnesium has been intercalated from the surface by the process of the diffusion. This physical adsorption and intercalation of Mg into B hexagonal layer lead to the formation of $MgB_2$ films [2]. However, this yields non-uniform formation of $MgB_2$ films. But in our electrochemical technique the process of intercalation is in-situ and uniform, homogeneous film would be obtained.

## IV. CONCLUSION

The low cost, lightweight thin films of MgB$_2$ were successfully synthesized first time by using novel electrochemical technique. The film deposited from the non-aqueous bath shows metallic nature and sharp superconducting transition is observed at 36.4 K. The sharp superconducting transition at 36.4 K indicates the device quality films and filamental type wire can be formed by this route.

## V. ACKNOWLEDGEMENTS

We would like to thank C-MET, Pune, India for providing the EDAX and SEM. Also we Wish to thank UGC, New Delhi for providing the financial support to carry out this work.

# FIGURE CAPTIONS

Figure 1  Polarization curve for the deposition of Magnesium- Boron film from (a) aqueous bath  (b) non-aqueous bath

Figure 2  Variation of cathodic current density with deposition time for the deposition of $MgB_2$

Figure 3  (a) X-ray diffraction pattern of $MgB_2$ film from aqueous bath onto Ag

(b) X-ray diffraction pattern of $MgB_2$ film from non- aqueous bath onto Ag

(c) X-ray diffraction pattern of $MgB_2$ from non aqueous bath onto stainless steel.

Figure 4  SEM photograph for the $MgB_2$ film deposited from the (a) aqueous and (b) non-aqueous bath

Figure 5  Temperature dependent electrical resistivity of $MgB_2$ films.

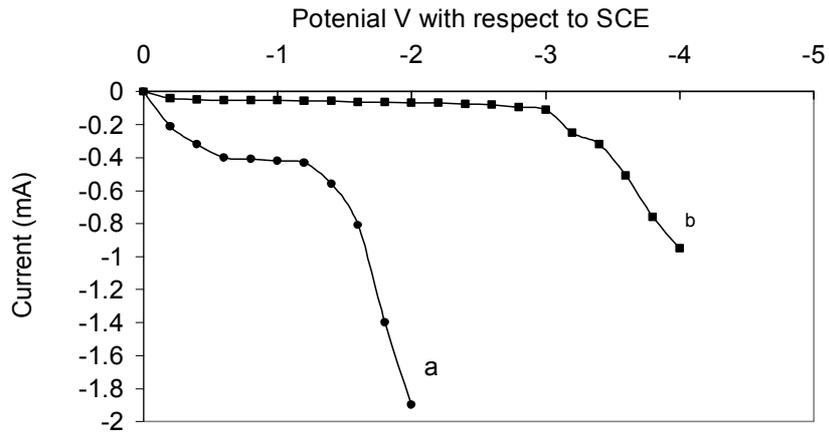

Fig.1 Polarization curve for the deposition of Magnesium- Boron film from
(a) aqueous bath     (b) non-aqoues bath

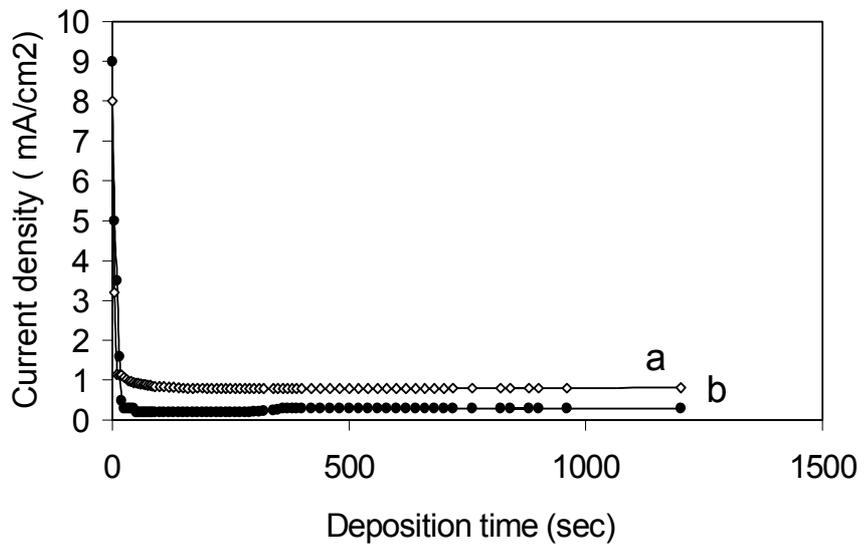

Fig. 2 Variation of cathodic current density with deposition time (a) aqueous (b) nono-aqueoes for the deposition of MgB2 onto Ag

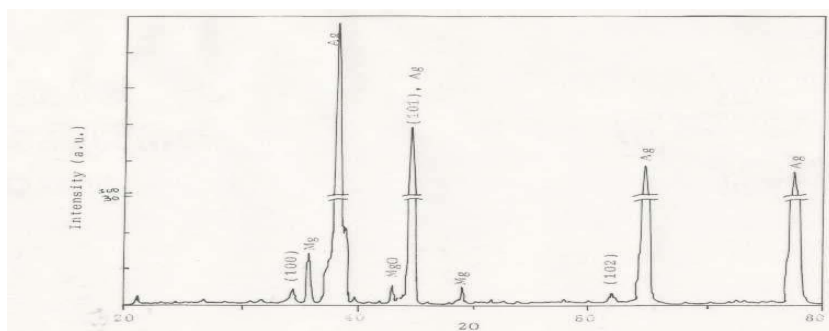

Figure 3 (a) X-ray diffraction pattern of MgB$_2$ film

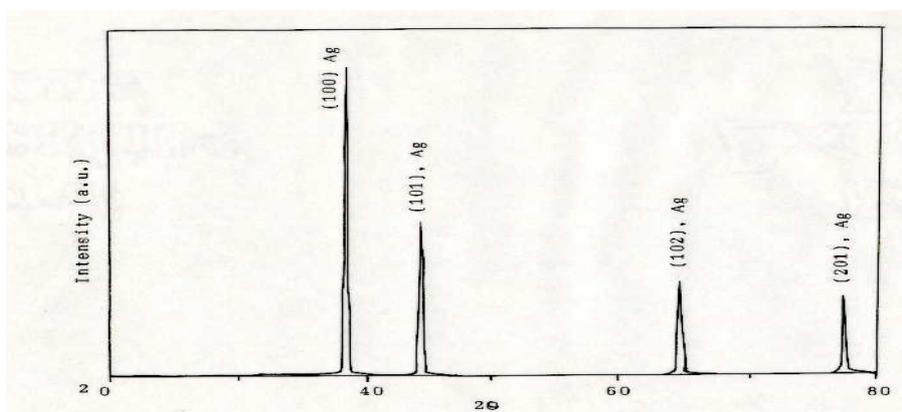

Fig 3 (b) X-ray diffraction pattern of MgB$_2$ film from

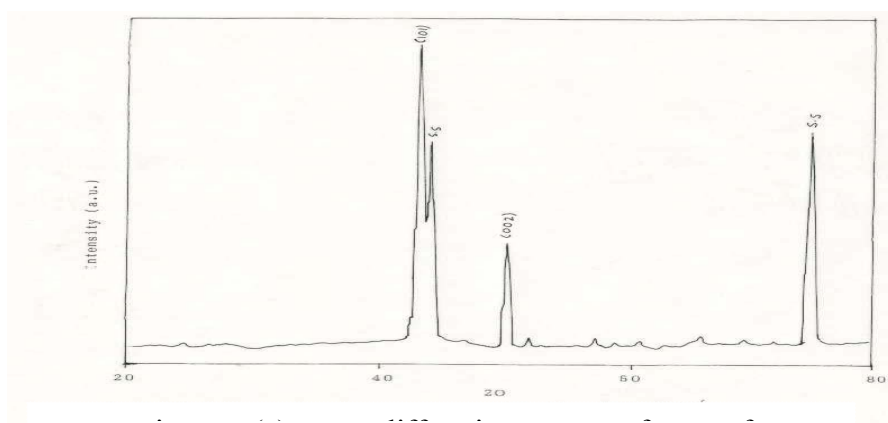

Fig 3 (c)

(a) 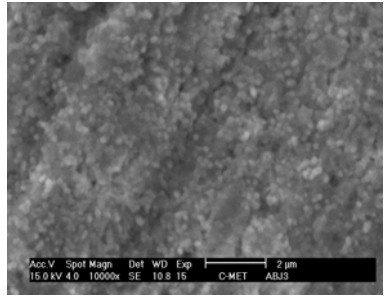 (b) 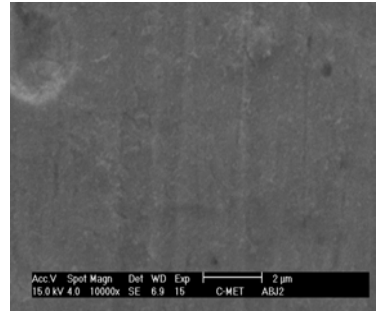

Fig. 4. SEM photograph for the MgB$_2$ film deposited

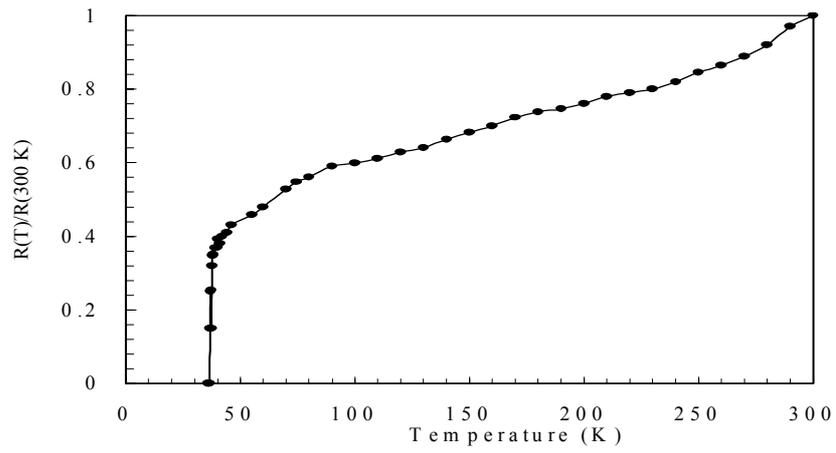

Fig.5. Temperature dependent electrical resistivity of MgB$_2$ films